\documentclass[prl,aps,twocolumn,superscriptaddress,notitlepage,longbibliography]{revtex4-2}

\usepackage{amsmath}
\usepackage{amssymb}
\usepackage[T1]{fontenc}

\usepackage[unicode=true,colorlinks=true,citecolor=blue,urlcolor=blue]{hyperref}

\usepackage{bm}
\usepackage{epsfig}
\usepackage{multirow}
\usepackage{soul}

\renewcommand {\Im}{\mathop\mathrm{Im}\nolimits}
\renewcommand {\Re}{\mathop\mathrm{Re}\nolimits}

\renewcommand {\phi}{{\varphi}}
\newcommand {\rmi}{{\rm i}}
\newcommand {\rmd}{{\rm d}}

\newcommand {\e}{{\rm e}}

\usepackage[normalem]{ulem}

\let\oldaddcontentsline\addcontentsline
\renewcommand{\addcontentsline}[3]{}%
\begin{document}

\title{Klein tunneling of the laser coherence}

\author{Konstantin Manannikov}
 
 \author{Sagie Gadasi }
\author{Ekaterina Mironova }
\author{Eran Bernstein }

\author{Andrei Poliakov}
 \author{Nir Davidson}
 
\author{Alexander N. Poddubny}%
\email{poddubny@weizmann.ac.il}

\affiliation{Department of Physics of Complex Systems, Weizmann Institute of Science, Israel, Rehovot 7610001}

\begin{abstract}
We study theoretically the lasing synchronization of the two arrays of lasers with the complex mode dispersion, separated by a spectrally detuned barrier.  We demonstrate that for lasing at the Dirac point, the synchronization persists for an order of magnitude higher barriers than in the arrays with a usual parabolic dispersion or a purely dissipative coupling.
We interpret this effect as the Klein tunneling of the laser coherence through the barrier. Our numerical findings are supported by an analysis of the delocalization of the linearized eigenmodes of the arrays, which enhances the synchronization.
\end{abstract}

\maketitle

Understanding whether multiple lasers can synchronize despite the inevitable noise, such as quenched disorder in laser resonant frequencies, is an essential problem both for basic nonlinear physics~\cite{strogatz2018nonlinear} and for practical applications \cite{beam_combining_rewiev}. For nearest-neighbor dissipative coupling between the lasers, the synchronization condition is determined by the competition between noise and the coupling strength \cite{Strogatz1988}. Synchronization can be enhanced when the coupling between the lasers has both dispersive and dissipative components and is characterized by a non-Hermitian effective Hamiltonian $H\ne H^\dag$\cite{Sakaguchi, Moroney1dcomplex, Moroney2dcomplex}. This complex coupling can now be realized in practice, for example, in a degenerate cavity laser setup~\cite{Arwas2022,Gadasi2022,mahler2024} or coupled polariton condensates~\cite{Lagoudakis_2017,Lagoudakis2021}.

Laser synchronization with complex coupling has been studied so far only for laser arrays with a simple mode dispersion \cite{Sakaguchi, Moroney1dcomplex, Moroney2dcomplex}.
More complex band structures, potentially useful for lasing applications, can be inspired by topological photonics, which uses mode dispersion as a tool to engineer edge states of light. 
Topological photonics has now matured from initial demonstrations of propagating electromagnetic edge states  \cite{Haldane2008a,Wang2009,hafezi2013c,rechtsman2013} to advanced studies of quantum~\cite{Barik2018,Yan2021}, nonlinear~\cite{Smirnova2020}, non-Hermitian\cite{Bergholtz2021}, and active  structures~\cite{Bahari2017,Kruk2018,Ota2020}.
Lasing of topological photonic edge states was demonstrated  \cite{Bahari2017} and later shown to be robust to disorder ~\cite{Segev2018b,Diko2021}. A Dirac band structure with tailored photon loss has been put forward to allow scalable single-mode surface-emitting lasing ~\cite{Contractor2022,Kant2024}.
\begin{figure}[b!]
\center{
\includegraphics[width =0.999\linewidth]{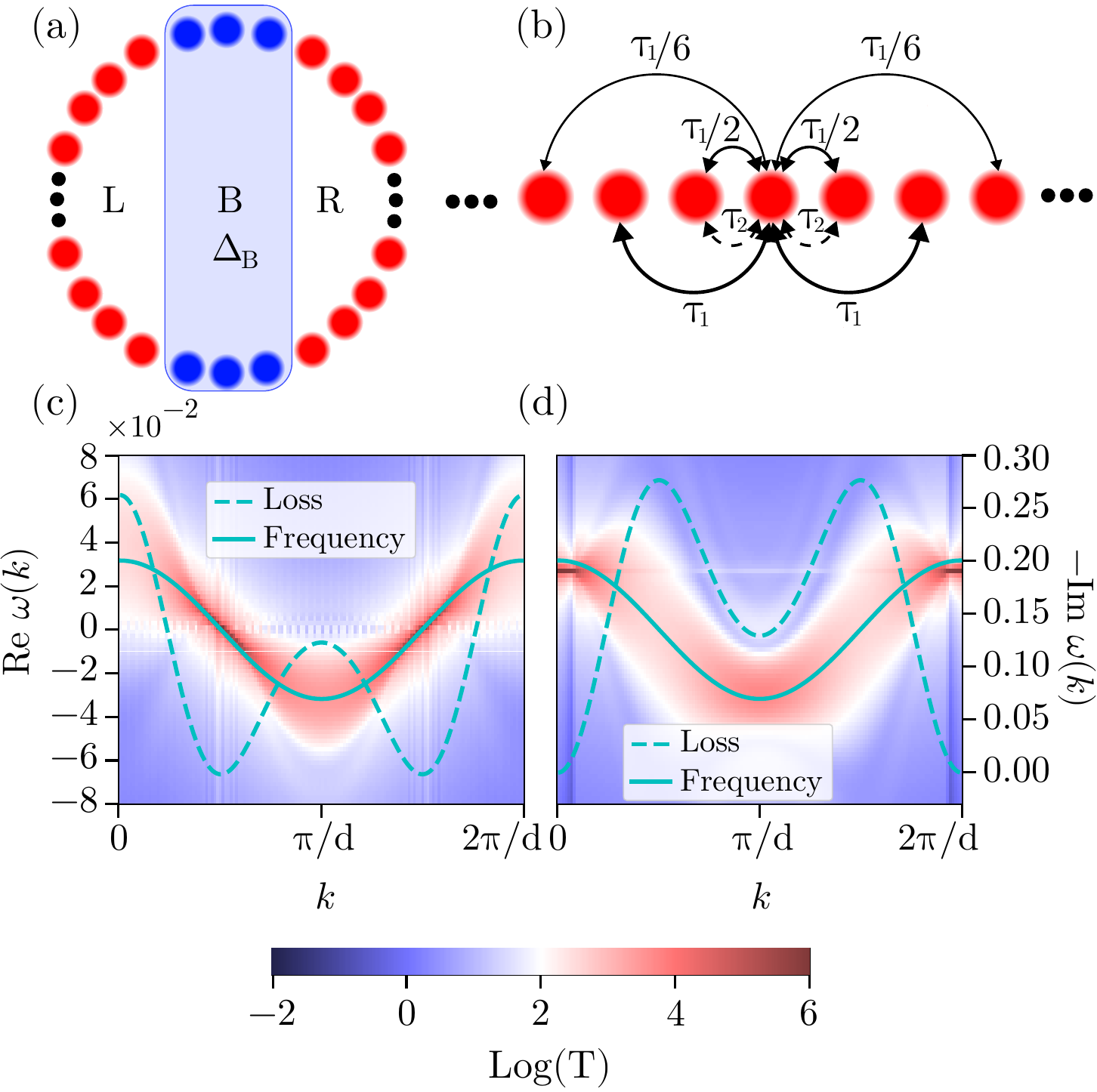}}
\caption{ 
(a) One-dimensional array of $N=100$ coupled lasers with periodic boundary condition (PBC) and double barrier composed of $N_B=3$ lasers with frequency detuning $\Delta_B$. $L$ and $R$ are the left and right regions of an array, respectively, and $B$ is the frequency detuned barrier that separates them. (b) A complex coupling scheme is implemented to achieve a minimal-loss lasing solution at k points with either linear or parabolic dispersion. Solid (dashed) lines denote dissipative (dispersive) coupling coefficients. (c-d) Lines: analytical calculation of band structure (solid) and loss landscape (dashed) obtained from diagonalizing the effective non-Hermitian Hamiltonian $H_{nn'}$. Depending on the sign of dissipative coupling, minimal loss states correspond to regions with linear dispersion ((c), $\tau_1<0$) or with parabolic dispersion ((d), $\tau_1>0$).
The color code represents the numerically calculated non-linear transmission $T$ vs wave vector $k$ and frequency $\omega$ on a logarithmic scale. The details are in text.}
\label{fig:linear-dispersion-1D}
\end{figure}

Despite this progress,  one phenomenon intrinsic to Dirac mode dispersion has not been investigated in the context of lasing -- chiral Klein tunneling~\cite{Katsnelson_2006,Katsnelson2020}. In Klein tunneling, originally proposed for relativistic quantum particles, particles with linear dispersion can cross a potential barrier with a tunneling amplitude that is independent of the barrier thickness. While photonic counterparts of  Klein tunneling have been recently studied ~\cite{MotiKlein,Ni2018,Solnyshkov2022,Nakatsugawa2024}, its impact on active lasing structures and their coherence remains unclear.

\begin{figure}[t!]
\center{
\includegraphics[width =0.99\linewidth]{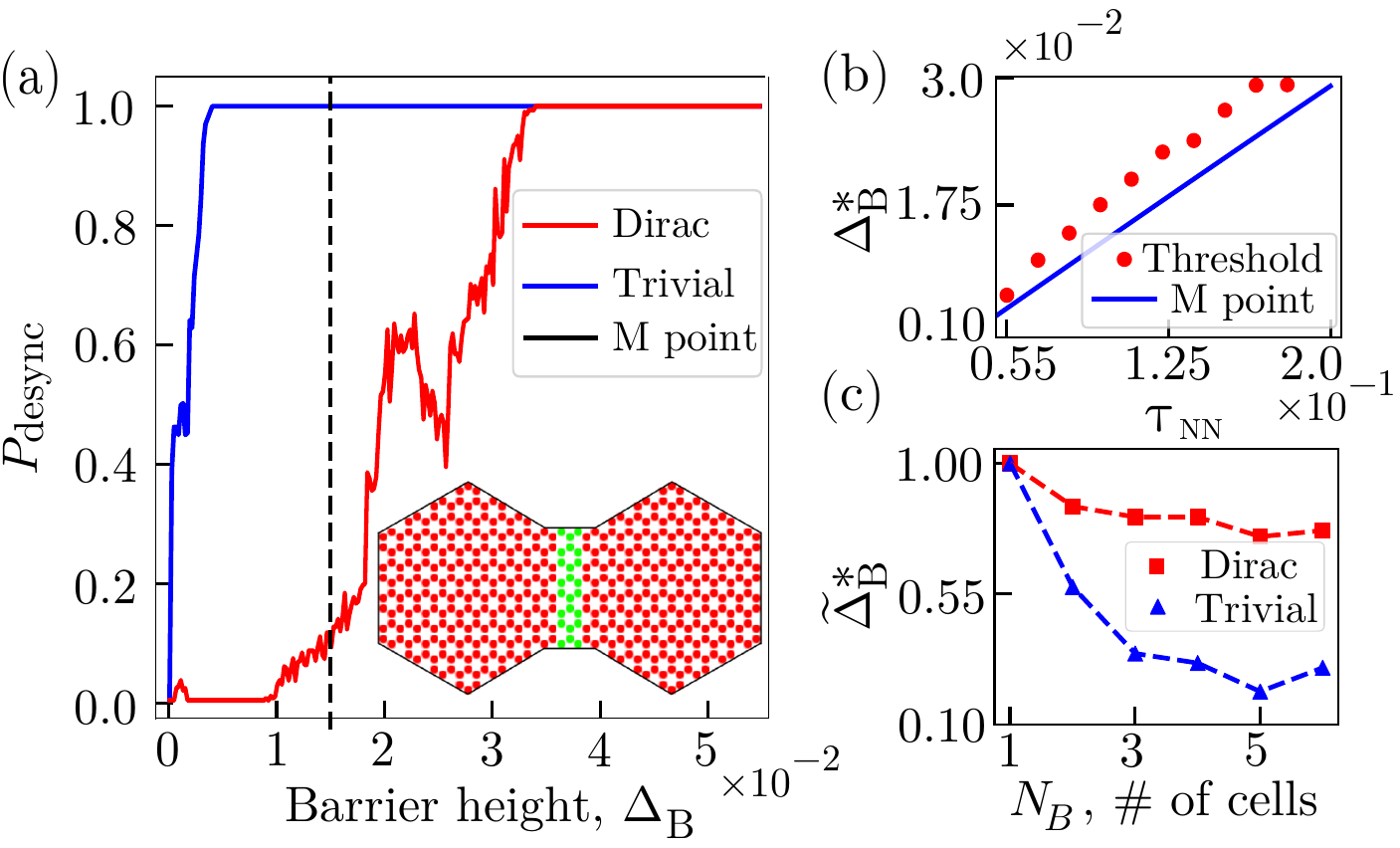}}
\caption{(a) Probability of desynchronization of the arrays left and right of the barrier for the trivial dispersion (blue dots) and for the Dirac dispersion (red dots). The dashed line indicates the position of the M  point for the Dirac band structure. The inset shows the geometry of the laser array with a barrier in the middle, where green lasers are detuned by $\Delta_B$. (b) Dependence of critical barrier height for synchronization $\Delta_B^*$ on the nearest neighbor coupling value $\tau_{\rm NN}$. The blue line shows the frequency of the $M$ point of the Brillouin zone.  (c) Dependence of the normalized critical barrier height $\tilde{\Delta}_B^* (N_B) = \Delta_B^* (N_B)/\Delta_B^* (1)$  on the width of the barrier, see Sec.~S2 of Ref.~\cite{Note1} for more details. }
\label{fig:sim 2D thr1}
\end{figure}

In this Letter, we theoretically investigate an analog of the Klein tunneling between two coupled laser arrays separated by a frequency-detuned barrier.
We consider a one-dimensional ring geometry [Fig.~\ref{fig:linear-dispersion-1D}(a)] and a two dimensional hexagonal lattice geometry [Fig.~\ref{fig:sim 2D thr1}(a)].
We tune the complex coupling between lasers such that both geometries lase at the $K$ point of the Brillouin zone with gapless Dirac (linear) dispersion.
We then use rigorous simulations of coupled laser dynamics \cite{Rogister2004} to demonstrate significant enhancement of synchronization for such linear dispersion, by tunneling through the unsynchronized barrier. Such tunneling-induced synchronization persists for much higher barriers than for coupling schemes with flat or parabolic dispersions. Moreover, the dependence of synchronization on barrier width is suppressed for linear dispersion, as expected in Klein tunneling.
We further support our Klein tunneling interpretation by showing that the enhanced synchronization is mediated via de-localized lowest loss modes of the linear
effective Hamiltonian. More calculation details are given in Sec.~S1 in Ref.~\cite{Note1}.


{\it Engineering complex dispersion in one dimension.}
We first demonstrate Klein tunneling of coherence for a gapless linear dispersion using a simple one-dimensional model of $N=100$ coupled lasers on a ring geometry with periodic boundary conditions, as shown in Fig.~\ref{fig:linear-dispersion-1D}(a). A double barrier $B$, composed of $N_B={3\div 21}$ lasers, separates the left region $L$ from the right region $R$. 
The barrier lasers are frequency detuned by a large enough $\Delta_B$ to prevent them from synchronizing with the lasers in regions $L$ and $R$. 

Each laser is coupled to its nearest, next-nearest, and next-next-nearest neighbors by the complex couplings $H_{n,n-1}=\tau_2+\rmi\tau_1/2$, $H_{n,n-2}=\rmi\tau_1$, and $H_{n,n-3}=\rmi\tau_1/6$, as shown in Fig.~\ref{fig:linear-dispersion-1D}(b). The parameters $\tau_1 = \pm 0.06$ and $\tau_2=  0.1 $ are the dissipative and dispersive couplings, respectively. The chosen ratio of the dissipative terms ensures that the sign of $\tau_1$  determines whether minimal loss lasing states occur at points with a linear [Fig.~\ref{fig:linear-dispersion-1D}(c)] or with a parabolic [Fig.~\ref{fig:linear-dispersion-1D}(d)] dispersion.  

The {\it linear} dynamics of the electric field  amplitudes $E_n$ is governed by a system of  equations
\begin{equation}\label{eq:linear}
    \frac{\rmd E_n}{\rmd t}=-\rmi H_{nn'}E_{n'}.\:
\end{equation}
Here, the effective non-Hermitian Hamiltonian matrix has nonzero elements $H_{n,n-1}$, $H_{n,n-2}$ and $H_{n,n-3}$ described above, and $H_{n,n}= \omega_n$, the detuning of a natural frequency of the lasers, where $\omega_n=0$ in regions $L$ and $R$ and $\omega_n=\Delta_B$ in the barrier region $B$ (in units of the free spectral range FSR=$2 \pi/T_r$, where $T_r=1$ is the cavity roundtrip time in our simulations, see Sec.~S1 in Ref.~\cite{Note1}). The system has linear reciprocity: the matrix is complex symmetric, $H_{nm}=H_{mn}$.  For periodic boundary conditions, the linear eigenstates of Eq.~\eqref{eq:linear} are $E_n(t)=\exp[\rmi kn-\rmi\omega(k)t]$, where $k$ is the Bloch wave vector and $\omega(k)$ is the complex eigenfrequency.
Solid and dashed curves in Fig.~\ref{fig:linear-dispersion-1D}(c,d) show the mode dispersion $\Re\omega(k)$ and mode loss $-\Im\omega(k)$ for negative (c) and positive (d) $\tau_1$.
$\Re\omega(k)=\tau_2\cos k$ and is linear around $k=\pm\pi/2$ and parabolic around $k = 0$. 

Importantly, due to nonlinear laser-mode competition, lasing occurs around the wavenumbers $k$ which minimizes the linear loss $-\Im\omega(k)$. For $\tau_1<0$ minimal loss is achieved at $k=\pm\pi/2$ yielding a lasing state with linear  dispersion (Fig.~\ref{fig:linear-dispersion-1D}(c)), while for  $\tau_1>0$  minimal loss is at $k = 0$ yielding a lasing state with parabolic  dispersion (Fig.~\ref{fig:linear-dispersion-1D}(d)). The color in Fig.~\ref{fig:linear-dispersion-1D}(c,d) encodes the logarithm of calculated effective non-linear transmission $T$ (see details in Sec.~S1 in \footnote{Online Supplementary Materials}). The bright blue color corresponds to large mode loss, in agreement with the linear calculation.

{\it Laser synchronization through a barrier.}

\begin{figure}[t!]
\center{
\includegraphics[width =0.99\linewidth]{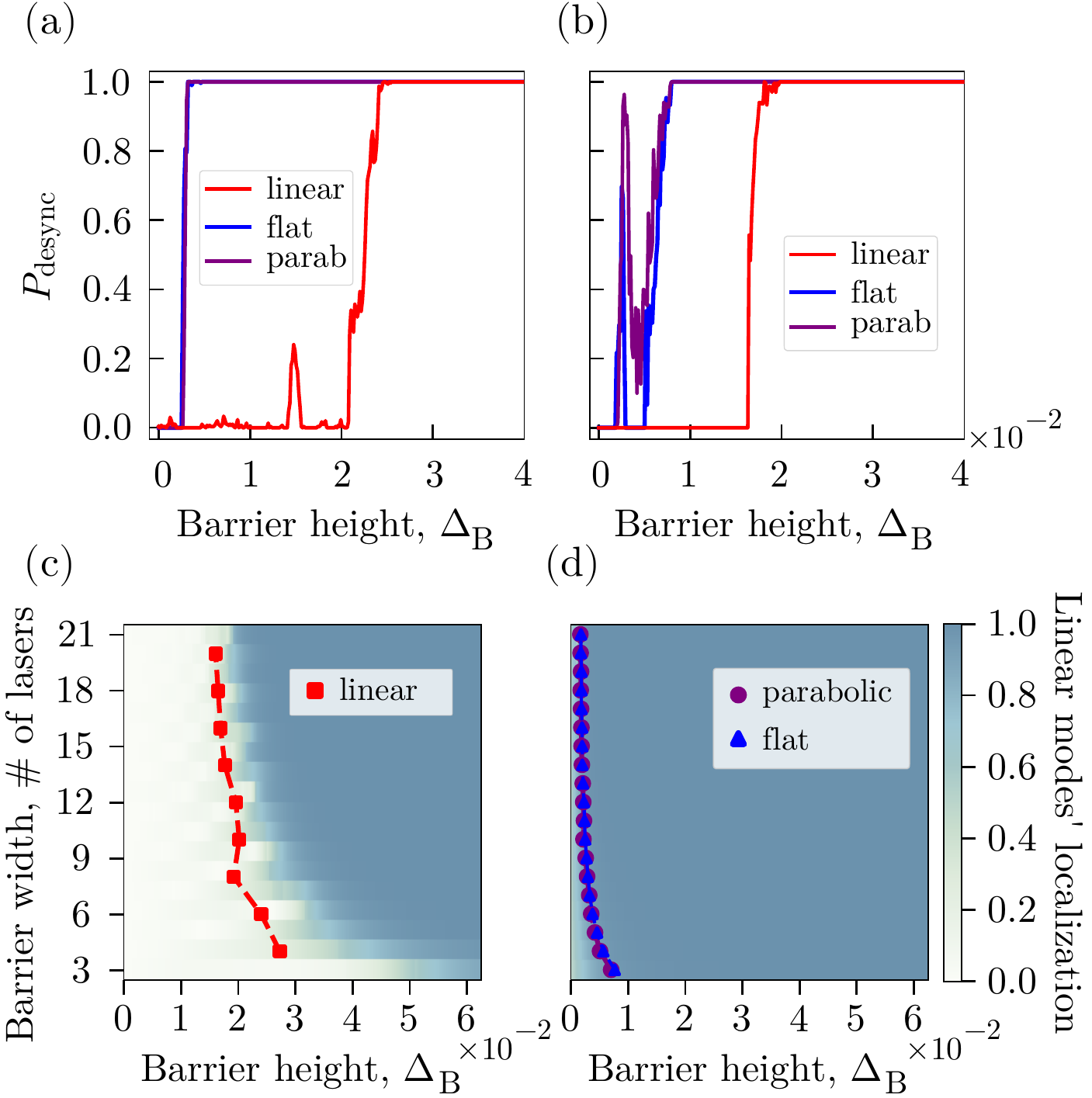}}
\caption{ (a) Probability of de-synchronization for barrier width $N_B = 10$ and different $\Delta_B$ for $\tau_1<0$ (linear dispersion), $\tau_1 > 0$ (parabolic dispersion) and $\tau_2 = 0$ (flat dispersion). (b) Same as (a), but for a reduced basis of only the $N_m = 4$ lowest-loss modes. (c-d) Color map represents the localization of the first 4 eigenmodes of the coupling matrix for different barrier heights and widths for  (c) $\tau_1 <0$ and (d) $\tau_1  >0$. Lines on color plots mark the desynchronization threshold, obtained from the non-linear simulation, and agree well with the linear localization transition.  The calculations were performed for  $N = 100$ coupled lasers with periodic boundary conditions, $p= 1.5$, and $\gamma = 0.1$. }
\label{fig:sim 1D thr1}
\end{figure}

In the absence of disorder, any finite dissipative coupling $\tau_1$ ensures that the lasers in each region of Fig.~\ref{fig:linear-dispersion-1D}(a) are always synchronized. However, lasers in region $L$ will synchronize with lasers in region $R$ only if the coherence can tunnel through the frequency-detuned barrier separating them.  
We study this coherence tunneling for gapless linear dispersion and for parabolic dispersion in search of an analog to the Klein tunneling effect.  

For this, we solved a system of {\it nonlinear} laser rate equations (LRE)  
$\rmi \dot E_n=(H_{nn'}+U_{nn'})E_{n'}$, where $H_{nn'}$ is the {\it linear} effective Hamiltonian of Eq.~\eqref{eq:linear}, and $U_{nn'}$ is a new {\it nonlinear} term
\begin{equation}\label{eq:non-lin model}
    U_{nn'} =  i \delta_{nn'}\Big[ \dfrac{p}{1 + |{E}_n|^2 / I_{\rm sat} } - \gamma \Big ] 
\end{equation}
that describes the saturable laser gain \cite{koechner2013solid}. The parameter $I_{sat}$ is the saturation intensity, $p$ is the pump rate, and $\gamma$ is the uniform constant loss. We used the values $p = 1.5, \; \gamma = 0.1, \; I_{\rm sat} = 1$ in the simulations. 
A small noise was also added to break the degeneracy between the left and right sides of the array ~\cite{noise_comment}. 

Results of the nonlinear simulation are shown in Fig.~\ref{fig:sim 1D thr1}(a-d). Figure~\ref{fig:sim 1D thr1}(a) presents the probability of desynchronization as a function of the barrier frequency height $\Delta_B$ for a fixed barrier width of $N_B=10$ lasers for both linear ($\tau_1 {<} 0$, red) and parabolic ($\tau_1 {>} 0$, purple) dispersion. 
The irregular shape of the curves in Fig.~\ref{fig:sim 1D thr1}(a,b) reflects the finite-sample estimate of $P_{\rm desync}$ over different initial conditions combined with a binary synchronization criterion. Since the detailed line shape of $P_{\rm desync}(\Delta_B)$ is not expected to be universal in a finite, strongly nonlinear system, we focus on the robust step-like crossover and the associated threshold scale.
Calculation details are given in Sec.~S2 of Ref.~\cite{Note1}.
As clearly seen,  when the barrier height crosses some threshold value, the lasers to the left and to the right of the barrier lose synchronization with each other. However, this threshold barrier height is sensitive to the dispersion law. The gapless linear dispersion manifested $\sim 10$ times higher threshold than the parabolic dispersion, indicating dramatically more robust laser synchronization. The threshold value for the linear dispersion is $\Delta_B^*\sim 2\times 10^{-2}$, where the barrier dispersion stops being linear ( Fig.~\ref{fig:linear-dispersion-1D}(c)). The flat dispersion case (blue, obtained for purely dissipative coupling, $\tau_2=0$) is nearly identical to the parabolic dispersion, in agreement with the Klein tunneling picture.

We repeated the LRE simulations of Fig.~\ref{fig:sim 1D thr1}(a) for barrier widths of $N_B= {3\div 21}$ lasers. We identified a sharp threshold of the barrier height leading to loss of synchronization between $L$ and $R$ lasers for all barrier widths and plotted them for linear dispersion (Fig.~\ref{fig:sim 1D thr1}(c), red) and for parabolic and flat dispersion (Fig.~\ref{fig:sim 1D thr1}(d), purple and blue, respectively). This desynchronization threshold is enhanced by an order of magnitude for linear dispersion across all barrier widths, demonstrating the generality of the Klein tunneling effect in our system.
Moreover, Fig.~\ref{fig:sim 1D thr1}(c)  shows that for wide barriers the critical barrier height for desynchronization $\Delta_B^*$ is nearly independent of the barrier width, as expected from the Klein tunneling~\cite{Katsnelson2020}. 

{\it Linear analysis.}
We further test the correspondence between the synchronization enhancement and the enhancement of the tunneling through the barrier by resorting to a linear analysis of the tunneling, corresponding to the localization of the lowest loss modes of the linear effective Hamiltonian $H$, characterized by the eigenvectors ${v_n}^{(i)}$ and eigenvalues 
 $\lambda^{(i)}$.

We start by decomposing electric field  ${E_n}(t)$ as 
\begin{equation}\label{eq:expansion}
  E_n(t) =  \sum_{i = 1}^{N_m} a_i(t) {v_n}^{(i)}
\end{equation}
Here, we considered only the first $N_m$ eigenvectors with a minimal loss $|\Im \lambda|$ in the decomposition.
Projecting  Eqs.~\eqref{eq:linear},\eqref{eq:non-lin model} onto the mode $i$ we find 
\begin{equation}\label{eq:LRE2}
  i \frac{ a_i(t)}{\rmd t} = \lambda_i a_i(t) + \sum_{j = 1}^{N_m}  a_j(t) \langle i|U|j\rangle\:.
\end{equation}
Here, we  took into account that  $H$ is a complex symmetric matrix and  defined a non-conjugate scalar product in the form of 
\begin{equation}
\begin{gathered}
  \langle i|j\rangle \equiv \sum\limits_{n=1}^Nv^{(i)}_n v^{(j)}_n \:,
\end{gathered}
\end{equation}
so that  for the eigenvectors of $H$ one has $\langle i|j\rangle= \delta_{ij}$.

Equation~\eqref{eq:LRE2} clearly indicates that if the linear eigenmodes are localized on different sides of the array, the matrix element $ \langle i|U|j\rangle= 0$ and so these modes are fully decoupled. Thus, there is no reason to expect a synchronization between them.

To demonstrate this explicitly, we repeat the nonlinear simulations of Fig.~\ref{fig:sim 1D thr1}(a) projected onto a reduced space of the $N_m=4$ lowest loss linear modes, as described in Sec.~S4 in Ref.~\cite{Note1}. The results, presented in Fig.~\ref{fig:sim 1D thr1}(b), are similar to Fig.~\ref{fig:sim 1D thr1}(a), verifying that the improved synchronization robustness for linear dispersion can indeed be captured in the reduced basis. Since only a few eigenmodes are retained, the reduced model should not be expected to reproduce the detailed line shape of the full LRE simulations. This truncation makes the curve in Fig.~\ref{fig:sim 1D thr1}(b) less regular, but the approximate step-like transition and its threshold scale are preserved. Next, we calculate the localization of these $4$ modes for different barrier heights and widths. 
Figures~\ref{fig:sim 1D thr1}(c) and (d) present color maps of the relative difference of the mode intensities $I_L$ and $I_R$, averaged over the left and right halves of the array, respectively. {Dark}  color corresponds to the uncoupled modes localized either on the left or on the right, and {bright} color corresponds to the delocalized modes. 
The transition from delocalized to localized modes occurs at nearly the same barrier width and height as the transition from synchronization to desynchronization in the nonlinear LRE simulations, further supporting our Klein tunneling interpretation: the delocalized modes mediate laser synchronization. 
We have also verified that the enhancement synchronization in the linear dispersion regime persists over a wide range of lasing parameters, such as the pump strength, see details in Sec.~S3 of Ref.~\cite{Note1}.

\begin{figure}[t!]\center{\includegraphics[width =0.99\linewidth]{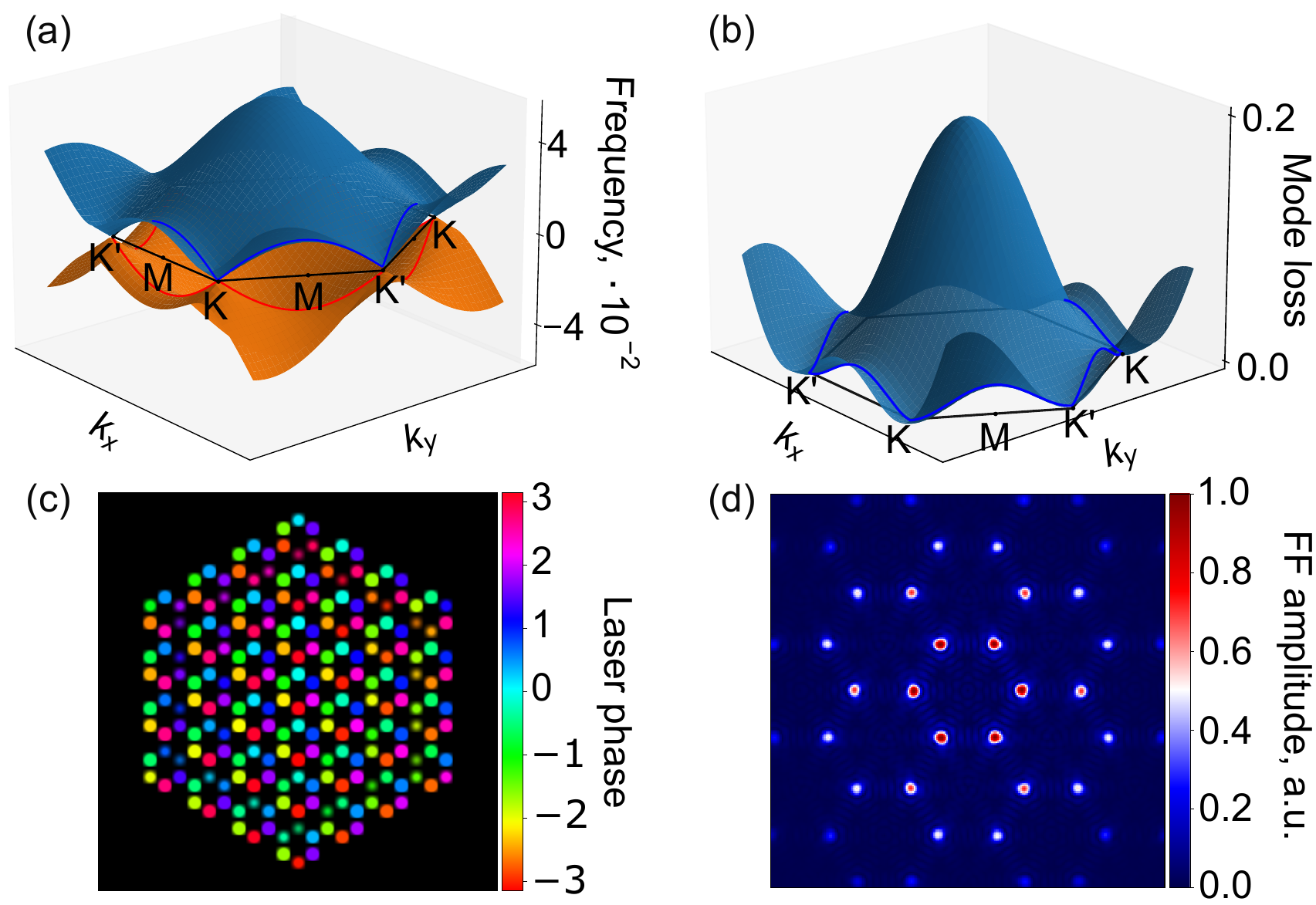}}
\caption{(a-b): Band structure of honeycomb laser array. Picture (a) shows the frequencies of two eigenvalues of the Bloch Hamiltonian, while the loss of these eigenvalues is on (b). The minimal loss solution is localized around $K$ and $K'$ points and has a linear dispersion relation. (c,d): 
Near and far field of the laser array with the linear dispersion, confirming lasing around the $K$ and $K'$ points and a chiral phase structure. }
\label{fig:band structure}
\end{figure}

{\it Klein tunneling of coherence in two dimensions.}  We now proceed to the two-dimensional honeycomb structure. Similar to the 1D model discussed above, we engineer lasing at the Dirac point by using the complex Dirac dispersion $\Re \omega(\bm k)$ with minimal loss $|\Im \omega(\bm k)|$ at the Dirac points.   This is realized for  the dispersive nearest-neighbor coupling and dissipative next-nearest-neighbor coupling  as described by the following effective tight-binding Hamiltonian
\begin{equation}\label{eq:H1}
   H(\bm k)
    \begin{pmatrix}
    E_A\\ E_B
    \end{pmatrix}
    =\omega(\bm k)\begin{pmatrix}E_A\\E_B
    \end{pmatrix}\:,\:\: 
     H(\bm k)=\begin{pmatrix}
    \tau_{\rm NNN} \widetilde g_{\bm k} &
    \tau_{\rm NN} g_{\bm k}\\
    \tau_{\rm NN} g_{-\bm k} & \tau_{\rm NNN} \widetilde g_{\bm k}
    \end{pmatrix}\:.
\end{equation}
with $\tau_{\rm NN} = 0.1$, $\tau_{\rm NNN} = -0.03\rmi $.  Here, $E_{A}$ and $E_B$ are the laser fields at the two sites $A$ and $B$ of the honeycomb unit cell,  $\bm k\equiv (k_x,k_y)$ is the Bloch wave vector,  
\begin{align}
    g_{\bm k}&=\e^{\rmi k_x}+2\e^{-\rmi k_x/2}\cos \frac{\sqrt{3}k_y}{2}\:,\\ \nonumber
    \widetilde g_{\bm k}&=2\cos \sqrt{3}k_y+4\cos\frac{3k_x}{2}\cos\frac{\sqrt{3}k_y}{2}\:.\nonumber
\end{align}

 Figure~\ref{fig:band structure}(a,b) shows the real (a) and imaginary parts (b) of
$\omega(\bm k)$ calculated  by diagonalizing  Eq.~\eqref{eq:H1}. The dispersion in  Fig.~\ref{fig:band structure}(a) manifests characteristic Dirac cones, and the loss in  Fig.~\ref{fig:band structure}(b) is minimized at the K points. Conversely, 
$\tau_{\rm NN}=0.1$ and $\tau_{\rm NNN}=0.03 \rmi$ yield lasing at the $\Gamma$ point of the Brillouin zone with parabolic (trivial) dispersion.
 Figure~\ref{fig:band structure}(c) shows the near field profile and 
Fig.~\ref{fig:band structure}(d) illustrates the far field, calculated as a Fourier transform of (c) in the form of  $E(\bm k)=\sum_{n}\e^{-\rmi\bm k\cdot \bm r_n}E_n/\sqrt{N}$. The far field has a characteristic hexagonal pattern with maxima at the $K$, $K'$ points.

Figure~\ref{fig:sim 2D thr1} 
 presents the simulations of the laser synchronization probability through the barrier. Similarly to the  1D case,  the left and right array sides (L and R) are separated by a barrier with frequency detuned by $\Delta_B$ as shown in the inset of Fig.~\ref{fig:sim 2D thr1}(a). 
Panel  (a)  shows how the synchronization parameter depends on the height of the barrier.  
For trivial dispersion, synchronization is destroyed already for small barrier heights $\sim 0.001$, while in the Dirac case synchronization persists for barriers that are higher by an order of magnitude, $\Delta_B\sim 0.01$. 

  Similar to the one-dimensional case of Fig.~\ref{fig:linear-dispersion-1D}(b), we have performed a detailed analysis of the linearized problem. This analysis, detailed in   Sec.~S4 in Ref.~\cite{Note1},  confirms that the underlying mechanism for synchronization is the delocalization of the lowest-loss eigenmodes. 
The lasing field profile is determined by the competition between the sample anisotropy and the nonlinearity. The anisotropy aims to split the modes with different symmetries inside the unit cell (which plays the role of the mode polarization), and the nonlinearity favors the chiral modes $p_x\pm\rmi p_y$, $d_{2xy}\pm \rmi d_{x^2-y^2}$ with the same intensity at all the lasers. An example of the near field is shown in Fig.~\ref{fig:band structure}(c) and Fig.~S3 in Ref.~\cite{Note1}. The synchronized laser mode has a chiral phase pattern with a vortex of charge 1 in each unit cell.

We have also studied the dependence of the tunneling on the nearest-neighbor coupling $\tau_{NN}$
 (Fig.~\ref{fig:sim 2D thr1}b).  
The results indicate that the threshold (red dots)  follows the position of the $M$ point, which has been independently found from the band structure and is shown by the blue line.  Indeed,  at such large detuning from the Dirac point, the dispersion stops being linear, and the  Klein tunneling is suppressed.  The position of the $M$ point, marked by the dashed vertical line in Fig.~\ref{fig:sim 2D thr1}(a), is also close to the desynchronization threshold. 
Finally,  Fig.~\ref{fig:sim 2D thr1}c shows that in the Dirac case, coherence tunneling persists up to large barrier widths, while in trivial cases, tunneling is quickly suppressed when the barrier width increases.

{\it Summary}. We demonstrated that frequency synchronization between two laser arrays, separated by a frequency-detuned barrier, is significantly enhanced when lasing at the gapless linear dispersion point (in 1D) and at the Dirac point (in 2D). The lasing synchronization persists as long as the eigenmodes of the linear system remain coupled through the barrier, resulting in Klein tunneling of the laser coherence. Our findings could open new avenues to investigate the impact of the complex photonic band structure on laser synchronization. An interesting future research direction is the interplay between disorder, artificial gauge fields, and spin-orbit interactions~\cite{mahler2024}. Our results could be experimentally verified using a degenerate cavity laser setup~\cite{Tradonsky_2021}.

\acknowledgments 

We are grateful to Michael Aizenman, Boris~L.~Altshuler, and {Geva Arwas}   for fruitful discussions.
 We acknowledge financial support  by the Minerva Foundation.
The work of A.N.P. has been also supported by research grants
from the Center for New Scientists and from the Center for
Scientific Excellence at the Weizmann Institute of Science, and by the Quantum Science and Technology Program of the
Israel Council for Higher Education.


\begin{thebibliography}{38}
\bibitem [{noi()}]{noise_comment}%
  \BibitemOpen
  \href@noop {} {}\bibinfo {note} {A small detuning equal to FSR/3000 was added
  to the left part of the array}\BibitemShut {NoStop}%
\end{thebibliography}
%

\setcounter{figure}{0}

\setcounter{equation}{0}

\newpage \clearpage 
\renewcommand{\thefigure}{S\arabic{figure}}
\setcounter{equation}{0}
\renewcommand{\theequation}{S\arabic{equation}}

\onecolumngrid
\begin{center}{\large{\bf Supplementary Materials}}
\end{center}
\twocolumngrid
\let\addcontentsline\oldaddcontentsline

\tableofcontents

\def\thesection{S\arabic{section}}

\def\thesubsection{S\arabic{section}.\arabic{subsection}}

\def\theequation{S\arabic{equation}}

\def\thefigure{S\arabic{figure}}

\def\thetable{S\arabic{table}}


\def\@seccntformat#1{\csname the#1\endcsname\quad}

\setcounter{secnumdepth}{3}

\makeatother
\section{Simulation of the spectral response}\label{sec:bandstructure}
In this section, we describe the numerical procedure used to simulate the 
 spectral response of the structure to the injected field with a given wave vector $k_0$. 
This response, depending on the frequency $\omega$ and the wave vector,  is shown by colors in Fig.~\ref{fig:linear-dispersion-1D}(c,d) and probes the complex band structure of the eigenmodes.

The simulation  can be summarized as follows:
\begin{enumerate}
\item We start by injecting the normalized field distribution with initial wave vector $k_0$:  \[
E_n(0) = \dfrac{1}{N} \e^{\rmi k_0 n}\:.\]
\item Next, we numerically solve the laser rate equations, Eqs. (1,2) in the main text. After one cavity round-trip time T (set to 1), the field will become 
\[E_n(T) =A_n(T)\e^{\rmi \varphi(n)} = \sum\limits_{k=0}^{2\pi/a} a_k(T) e^{i k n}\:.\] 
To obtain the amplitude  $a_{k_0}(t)$ we  calculate the overlap: 
\begin{multline}\langle E_n(T), E_n(0)\rangle  \equiv  \sum\limits_{n=0}^{N} \dfrac{1}{N}e^{i k_0 n} \sum\limits_{k=0}^{2\pi/a} a_k^* (T) e^{-i k n}\\ = \frac{1}{N} \sum\limits_{k=0}^{2\pi/a} a_k^* (T) \sum\limits_{n=0}^{N}  \e^{\rmi k_0 n -\rmi k n} \approx \sum\limits_{k=0}^{2\pi/a} a_k^* ( T) \delta_{k_0,k} \\=  a_{k_0}^*( T)\:.
\end{multline}

The last step is to normalize  $a_{k_0}^*(T)$ by $\sqrt{\langle E_n(T), E_n(T)\rangle}$\:.

\item Finally, we make a Fourier transform from the set of \{$ a_{k_0}(0),  a_{k_0}(T),  a_{k_0}(2T), ...$\}. The resulting array will be \{$ a_{k_0}(\nu = 0),  a_{k_0}(\nu = \nu_0),  a_{k_0}(\nu = 2 \nu_0), ...$\}. The real part of the disperson curve is given by $\ln(a_{k_0}(\nu))$ and the transmission spectrum is defined as $|a_{\tau_0}(\nu)|$.
\end{enumerate}
We repeat this procedure for all the input wave vectors $k_0$.

\section{Probability of desynchronization and critical barrier height $\Delta_B^*$}
In this section, we describe the procedure for quantifying frequency synchronization in numerical calculations. We take a laser with the index $i$ from the right side of the array and another one with the index $j$ from the left side, and denote their phase difference as $\Delta_{ij}$. We are interested in their phase fluctuations over time (practically over the 1000 cavity roundtrip times) $\sigma_t[\Delta_{ij}]$. We denote the results of averaging over different pairs (i, j) as $\langle\sigma_t[\Delta_{ij}]\rangle_{\rm pairs}$. We define the lasers to be  synchronized if $\langle\sigma_t[\Delta_{ij}]\rangle_{\rm pairs} < 0.2$. To calculate the probability of desynchronization $P_{\rm desync}$, we average over different random initial conditions (typically around 200 realizations). To calculate critical barrier height $\Delta_B^*$ we first fix some initial barrier height $\Delta_B$ and calculate $\langle\sigma_t[\Delta_{ij}]\rangle_{\rm pairs, init\_cond} $. If this value is above 0.2, it means that $\Delta_B > \Delta_B^* $, and vice versa. Then we perform a binary search while $|\langle\sigma_t[\Delta_{ij}]\rangle_{\rm pairs,
 init\_cond}-0.2| > 0.02$.

\section{The role of pump strength}
In this section, we discuss that the laser synchronization enhancement is independent of the pump strength. We concentrate on the one-dimensional model described in the main text.

\begin{figure}[t!]

\center{
\includegraphics[width =0.75\linewidth]{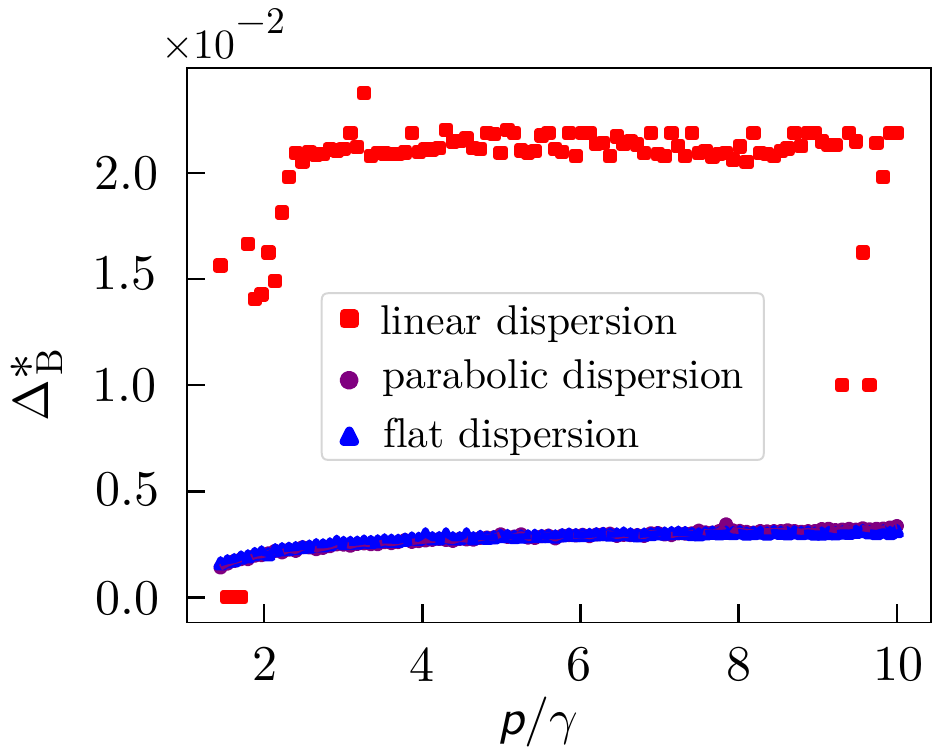}
}
\caption{Dependence of critical barrier height $\Delta_{B}^*$ on the pump strength in the 1D case. Flat dispersion and parabolic dispersion cases grow slowly with increasing pump, while the linear dispersion case saturates around $p/\gamma \approx 2.4$ and stays constant. The calculation has been performed for $\gamma=0.1$ and other parameters the same as in Fig.~\ref{fig:sim 1D thr1}.  }\label{fig:pump}
\end{figure}

Figure~\ref{fig:pump} shows how the tunneling threshold depends on the pump strength for a fixed barrier width equal to 10 lasers, $I_{sat} = 1$. Different-colored curves represent the coupling schemes used in the main text: linear, parabolic, and flat dispersion. 
When calculating Fig.~\ref{fig:pump}, we have shifted the matrix $H$ by a constant in such a way that the loss has been equal to  $\gamma = 0.1$. This procedure is required to reach the same minimal loss value for all parameters to maintain self-consistency.

The results in Fig.~\ref{fig:pump} demonstrate that for the chosen coupling parameters in the 1D model, the critical barrier height, when the synchronization is destroyed, is always higher for the model with the linear dispersion, independent of the pump strength. The  2D  is more sensitive to the pump level. The best synchronization is achieved for $p/\gamma$ in the range approximately $1.3$ to $1.6$.

\begin{figure}[b!]
\center{
\includegraphics[width =0.75\linewidth]{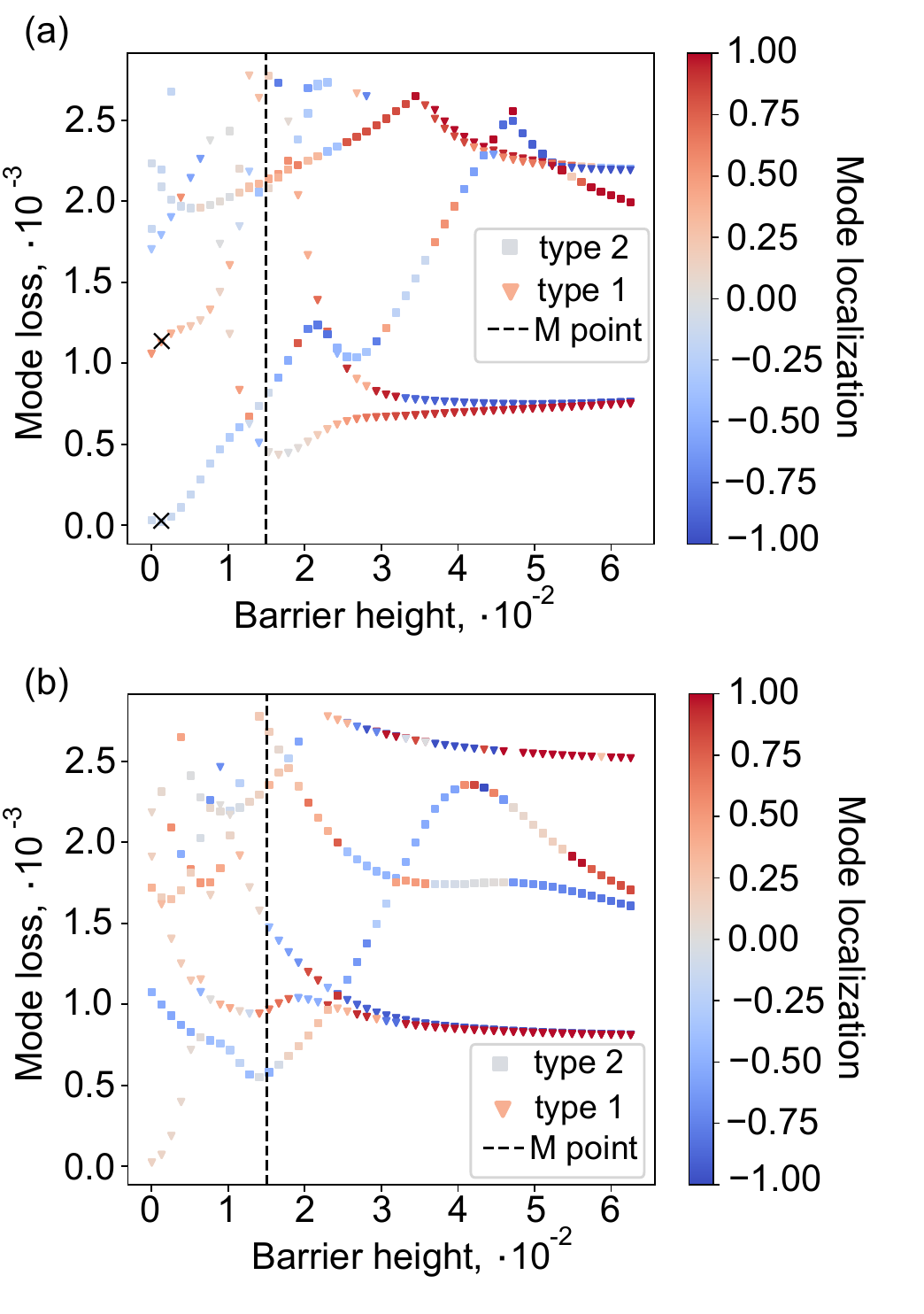}
}
\caption{Losses of first eight eigenmodes over different barrier heights for the Dirac case with (a) positive and (b) negative frequency. The type of the symbol corresponds to the type of the mode, as described in the legend, and the symbol color illustrates the mode localization.   The dashed vertical line shows the barrier height corresponding to the $M$ point. The calculation has been done in the linearized regime.}
\label{Dirac coupling lin analysis}
\end{figure}

\section{Linear modes localization for a  graphene lattice}

We now describe the relationship between the linear system eigenmodes and the lasing eigenmodes in the two-dimensional structure. Such analysis is significantly more involved than in the one-dimensional model with the linear dispersion for the following three reasons. The first reason is the fourfold polarization- and valley-degeneracy of the modes at the Dirac point. We designate these four degenerate modes as  \{$p_{x}, p_{y}, d_{x^2 - y^2}, d_{2xy}$\}. Formally, the distribution in the hexagonal unit cell with the hexagon vertices at the  sites $\bm r_n\equiv (x_n,y_n)$ for the four modes  can be obtained as 
\begin{align}\label{eq:modes}
    &E_{x}(\bm r_n)\propto x_n, &&E_{y}\propto y_n\:, \\
    &E_{2xy}(\bm r_n)\propto 2x_ny_n, &&E_{x^2-y^2}\propto x_n^2-y_n^2\:\nonumber,
\end{align}
 where the subscript denotes the mode symmetry.
The second complication arises from the system's nonlinearity. In particular, the modes Eq.~\eqref{eq:modes} are not homogeneous in space, while the lasing nonlinearity favors the lasers to have the same amplitude~\cite{Gadasi2022}.  These are not the modes Eq.~\eqref{eq:modes}, but their  linear combinations \{$p_{\pm}\equiv p_x\pm \rmi p_y, d_{\pm}\pm d_{2xy}\pm d_{x^2-y^2}$\}.  However, there is a third complication:  the finite structure considered in the simulation has a lower $C_{2v}$ symmetry  (see inset in Fig.~\ref{fig:sim 2D thr1} in the main text)~\cite{dresselhaus2007group}, which is lower than the infinite array. Due to the lowering of the symmetry, both pairs of modes $p_x$ and $p_y$ and $d_{2xy},d_{x^2-y^2}$ split.
 
Thus, the nonlinearity favors the symmetric modes 
\{$p_{\pm}, d_{\pm}$\} while the anisotropy favors the polarization-split modes. It is this competition that determines the complicated profile of the lasing mode.

We will now analyze this in more detail for varying barrier heights. We start by grouping the four modes by their far-field patterns. We define type-1 modes as \{$p_{x}, d_{2xy}$\}, and and type-2 modes as \{$p_{y}, d_{x^2 - y^2}$\}. In order to build a good approximation for a nonlinear lasing mode that is spatially homogeneous inside the unit cell, we take a linear combination of a type-1 mode and a type-2 mode with a prefactor of $1\rmi$. Such a linear combination makes sense only if the two modes have approximately the same frequency.

\begin{figure}[t!]
\center{
\includegraphics[width =0.55\linewidth]{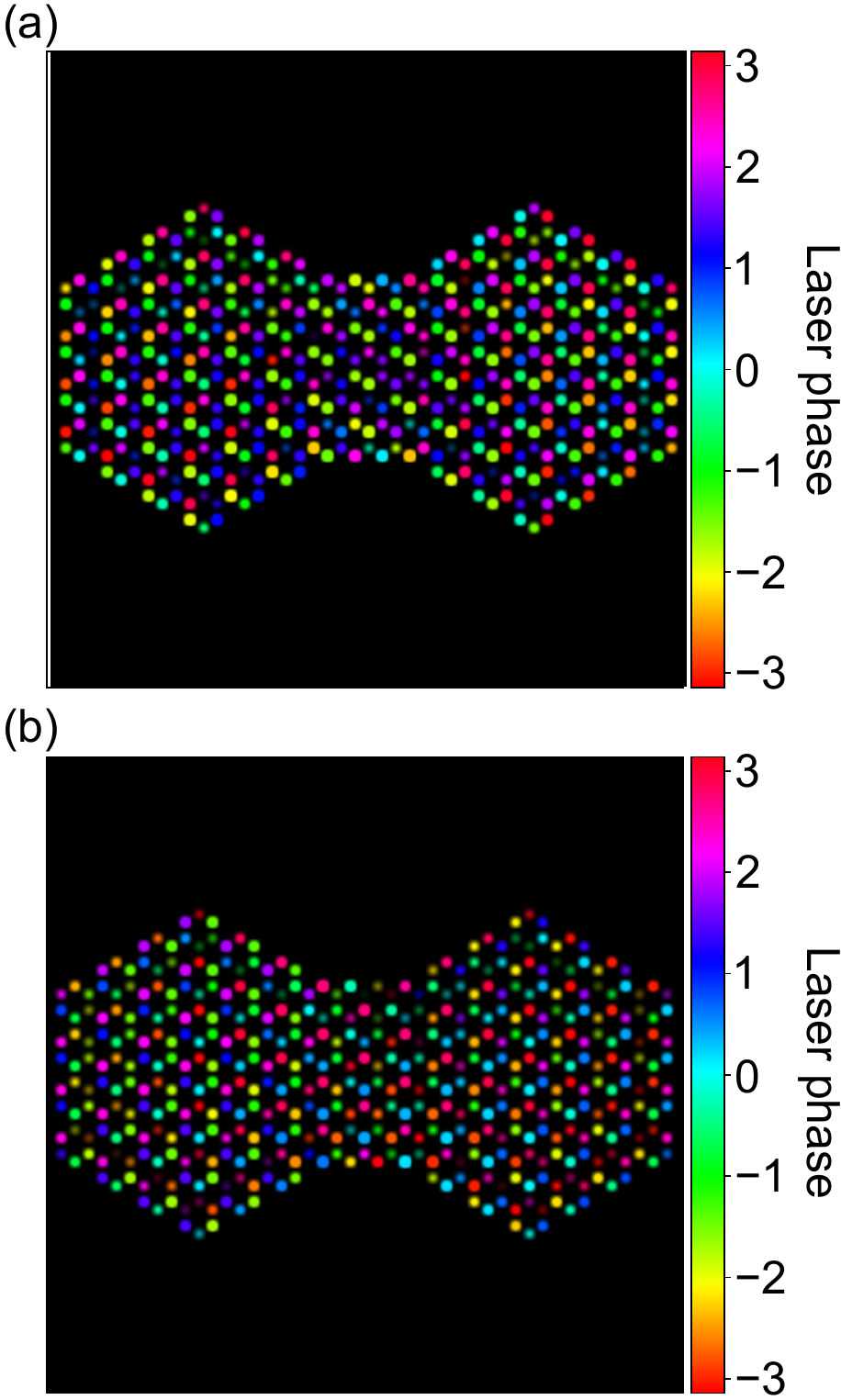}
}
\caption{Panel (a) demonstrates one of the possible results of LRE simulation with Dirac coupling. Panel (b) represents the most similar combination of linear eigenmodes with nearly equal frequency.}
\label{fig:example Dirac coupling lin analysis}
\end{figure}

In Fig.~\ref{Dirac coupling lin analysis}(a), we show the first four lowest-loss modes with the positive frequency, while the bottom picture demonstrates the first four lower-loss modes with the negative frequency.  We have verified that, when synchronization is possible, the lasing solution is well approximated by a linear combination of two modes, types 1 and 2, with either positive or negative frequencies.
An example of such an approximation is illustrated in Fig.~\ref{fig:example Dirac coupling lin analysis}. The panels (a) and (b) present the numerically calculated mode phase profile and its approximation by a combination of two modes, indicated by the $\times$ symbol in Fig.~\ref{Dirac coupling lin analysis}.

As it can be seen from Fig.~\ref{Dirac coupling lin analysis}, for barrier height $\in [0, 1.5\cdot 10^{-2}]$ linear modes are very well delocalized, for barrier height $> 4\cdot 10^{-2}$ modes are localized, and $[1.5\cdot 10^{-2}, 4\cdot 10^{-2}]$ is an intermediate region, where modes are localized/delocalized partially. This localization's behavior is consistent with the probability of desynchronization in the non-linear regime in Fig.~\ref{fig:sim 2D thr1}(a) from the main text.

\end{document}